\documentstyle[aps,prl,epsf,floats]{revtex}
\draft
\begin{document}
\def\SNG{{\em Physical Review Style and Notation Guide}}
\def\LUG {{\em \LaTeX{} User's Guide \& Reference Manual}}
\def\btt#1{{\tt$\backslash$\string#1}}%
\def\REVTeX{REV\TeX}
\def\AmS{{\protect\the\textfont2
        A\kern-.1667em\lower.5ex\hbox{M}\kern-.125emS}}
\def\AmSLaTeX{\AmS-\LaTeX}
\def\BibTeX{\rm B{\sc ib}\TeX}
\twocolumn[\hsize\textwidth\columnwidth\hsize\csname@twocolumnfalse%
\endcsname
\title{Density expansion for transport coefficients: Long-wavelength versus
       Fermi surface nonanalyticities}
\author{F. Evers, D.Belitz and Wansoo Park\cite{byline}}
\address{Department of Physics and Materials Science Institute,\\ 
 University of Oregon,\\
 Eugene, OR 97403}
    
\date{\today}
\maketitle
\begin{abstract}
The expansion of the conductivity in $2$-$d$ quantum Lorentz models in terms of
the scatterer density $n$ is considered. We show that nonanalyticities in the 
density expansion due to scattering processes with small and large
momentum transfers, respectively, have different functional forms.
Some of the latter are not logarithmic, but rather of power-law nature,
in sharp contrast to the $3$-$d$ case. In a $2$-$d$ model with 
point-like scatterers we find that the leading nonanalytic correction to the 
Boltzmann conductivity, apart from the frequency dependent weak-localization 
term, is of order $n^{3/2}$.
\end{abstract}   
]
It is well known from the statistical mechanics of fluids that for 
transport coefficients, as opposed to thermodynamic
quantities, no virial or density expansion exists\cite{Peierls}. 
Let us consider a classical Lorentz gas\cite{Hauge},
i.e. a single particle moving in a static array
of scatterers with scatterer density $n$, as a simple model of a classical 
fluid. In such a system in three dimensions ($d=3$), the diffusion coefficient,
$D$, as a function of $n$ has the form,
\begin{mathletters}
\label{eqs:1}
\begin{equation}
D/D_{\,\rm B} = 1 + D_1 n + D_{2\ln}n^2\ln n + D_2 n^2 + o(n^2)\ .
\label{eq:1a}
\end{equation}
Here $D_{\,\rm B}$ denotes the Boltzmann diffusivity, $D_1$,
$D_{\rm 2\ln}$, and $D_2$ are numbers, and $o(n^2)$
denotes terms that vanish faster than $n^2$ for $n\rightarrow 0$. In
$2$-$d$ systems, a similar nonanalyticity appears, but at one lower
order in the density expansion,
\begin{equation}
D/D_{\,\rm B} = 1 + D_{1\ln}n\ln n + D_1 n + o(n)\quad.
\label{eq:1b}
\end{equation}
\end{mathletters}%

The nonanalyticities in Eqs.\ (\ref{eqs:1}) are not specific to the Lorentz
models, but are also present in real fluids\cite{2dfootnote}. 
They are a result of long range 
dynamical correlations in the system. If one performs a cluster expansion
in a $d$-dimensional system, then ring collisions, i.e. processes where the
scattered particle collides with a scatterer to which it returns after
having scattered off a number of other scattering centers, lead to
a logarithmic infinity in the density expansion at order 
$n^{d-1}$\cite{Fox}. This
divergence is cut off by the mean-free path. Through the density dependence
of the latter, this translates into a logarithmic nonanalyticity at that
order. The nature of the higher order terms is not known, but they are
believed to also contain logarithmic nonanalyticities.

Since the mechanism that is believed to lead to these effects in classical
systems is rather general,
one would expect a similar effect to occur in the transport coefficients
of quantum mechanical particles. This is indeed the 
case\cite{LangerNeal}. Although the leading
expansion parameter is different\cite{KirkpatrickDorfman}, 
and performing the classical limit to make contact with the classical 
ring collisions is nontrivial\cite{ResiboisVelarde}, one
obtains for a $3$-$d$ quantum system again an expansion of the form given
in Eq.\ (\ref{eq:1a}). Specifically,
all of the coefficients in Eq.\ (\ref{eq:1a}) have been calculated for
a quantum Lorentz model that is a good representation of electrons injected
into Helium gas\cite{Wysolong}. 
It is apparent from the details of these calculations,
although not from the result, that in a quantum system there are two
physically distinct sources of the logarithmic nonanalyticity. These are,
(1) long-wavelength contributions, i.e. those dominated by scattering processes
with a momentum transfer on the order of an inverse mean-free path, 
and (2) Fermi surface contributions, which
are dominated by momentum transfers close to $2k_{\rm F}$, with $k_{\rm F}$ 
being the Fermi momentum. In a dilute system, these two length scales are
well separated.
Again, the nature of the higher terms is not known, but inspection of some
individual terms makes it appear likely that the next term in 
Eq.\ (\ref{eq:1a}) is of the form $n^3 (\ln n)^2$.

In this Letter we show that the fact that both of the mechanisms mentioned
above yield a logarithmic nonanalyticity is characteristic of $d=3$, and that
in $d=2$ some of the Fermi surface contributions lead to a 
{\em power law} nonanalyticity of the form $n^{3/2}$. Moreover, we
find that in a $2$-$d$ quantum Lorentz model with point-like scatterers,
the leading $n\ln n$ nonanalyticities cancel, so that the $n^{3/2}$ term
is the {\em leading} nonanalyticity, apart from the frequency dependent
weak-localization logarithm that appears in a $2$-$d$ quantum system. The
density expansion for the frequency dependent conductivity, $\sigma(\omega)$, 
(which in a quantum system is easier to calculate than the 
diffusivity) in such a model thus takes the form,
\begin{mathletters}
\label{eqs:2}
\begin{eqnarray}
    {\rm Re}\,\sigma(\omega)/\sigma_{\rm B} &=& 1  
      + \frac{2}{\pi}\frac{\gamma}{2\epsilon} \ln(\omega\tau)   
      + \sigma_{1\ln}\frac{\gamma}{2\epsilon} \ln\left(\frac{\gamma}
        {2\epsilon}\right) + \sigma_{1}\frac{\gamma}{2\epsilon}
    \nonumber\\
    && +\ \sigma_{3/2}    \left(\frac{\gamma}{2\epsilon}\right)^{3/2}
       + o\left(\frac{\gamma}{2\epsilon}\right)^{3/2}\quad,
\label{eq:2a}
\end{eqnarray}
where we have left out contributions that vanish as $\omega\rightarrow 0$.
Here $\sigma_{\rm B}{=}e^2\epsilon\tau/(2 \pi m)$ denotes the Boltzmann
conductivity, $\tau$ is the scattering mean-free time, $m$ is the electron
mass, $\epsilon=k_{\rm F}^2$, and $\gamma=m/\tau=n\sigma_T k_F$, with
$\sigma_T$ the total s-wave cross-section, is a convenient
expansion parameter that is proportional to the impurity density. 
The $\ln (\omega\tau)$ term is the so-called weak-localization correction 
(see below), whose prefactor has been known for some time. 
For the remaining coefficients
in Eq.\ (\ref{eq:2a}) we find, 
\begin{equation}
 \sigma_{1\ln} = 0\ ,\ \sigma_{1} = \frac{2}{\pi}\,(1-\ln 2)\ ,\ 
 \sigma_{3/2} = 3/2\sqrt{2}\ .
\label{eq:2b}
\end{equation}
\end{mathletters}%

Equations (\ref{eqs:2}) constitute our result. Before we sketch its
derivation, let us explain the physical origin of the
$\gamma^{3/2}$ term in Eq.\ (\ref{eq:2a}). In a degenerate system of
noninteracting fermions, the Pauli principle restricts the phase space 
that is available in scattering processes. These restrictions
lead, e.g., to the well-known
nonanalyticity of the Lindhard function at a momentum $q=2k_{\rm F}$.
The nature of this nonanalyticity is dimensionality dependent; it is
logarithmic in $d=3$, and a square root in $d=2$.
The same phase space restrictions lead to related nonanalyticities
in the scattering cross section, and hence in the transport coefficients.
This is precisely what we find.
The effect we predict is thus a consequence of the sharpness of the Fermi
surface. It is related to other phenomena resulting from the degenerate
nature of a Fermi gas at $T=0$, like, e.g., the Friedel oscillations, and
the Kohn anomaly.

We now outline the derivation of our results. Let us consider the standard
Edwards model of noninteracting electrons in $d=2$ in an environment of static,
spatially random scatterers. The model Hamiltonian reads,
\begin{equation}
H = \sum_{\bf k}\,(\epsilon_{{\bf k}}-\mu)\ a^{\dag}_{\bf k}a_{\bf k}
    + \sum_{{\bf k}, {\bf q}}\, $V({\bf q})$\ a^{\dag}_{{\bf k} + {\bf q}/2}\,
                                            a_{{\bf k} - {\bf q}/2}\quad,
\label{eq:3}
\end{equation}
where $a^{\dag}_{\bf k}$ and $a_{\bf k}$ denote the creation and 
annihilation operators, respectively, for electrons with wave vector ${\bf k}$,
$\mu$ is the chemical potential, $V({\bf q})$ is the Fourier transform 
of the electron-impurity scattering potential, and 
$\epsilon_{{\bf k}}={\bf k}^2/2m$. Throughout this paper we use units
such that $\hbar=1$. Also, we will be working at zero temperature, so we put
$\mu = \epsilon_{\rm F} = k_{\rm F}^2/2m$. Since we are dealing with 
noninteracting electrons, spin just leads to trivial factors of two and 
can be omitted. Standard diagrammatic perturbation theory\cite{Mahan} is 
formulated in terms of retarded (R) and
advanced (A) zero temperature Green's functions,
\begin{mathletters}
\label{eqs:4}
\begin{equation}
{\cal G}^{R,A}_{{\bf k},{\bf p}} (\omega) = \left\langle {\bf k} \left\vert
   \frac{1}{\omega - H \pm i0} \right\vert{\bf p}\right\rangle
\label{eq:4a}
\end{equation}
and their impurity averaged counterparts 
$\{ {\cal G}^{R,A}_{\bf k,p}(\omega)\}_{\rm dis}$, 
where $\{\ldots\}_{\rm dis}$
denotes the average over the quenched disorder. It is most convenient to
keep only the lowest order contribution to the self energy in the averaged
Green's function, so we will use as the building blocks of our perturbation
theory the following approximation to 
$\{ {\cal G}^{R,A}_{\bf k,p}(\omega)\}_{\rm dis}$,
\begin{equation}
G^{R,A}_{\bf k} = \frac{1}{\omega - (\epsilon_{\bf k} - \epsilon_{\rm F})
                  \pm i\gamma/2m}\quad.
\label{eq:4b}
\end{equation}
\end{mathletters}%
In writing Eq.\ (\ref{eq:4b}), we have specialized to the case of pure
s-wave scattering, or a point-like impurity potential, i.e. we have put
$V({\bf q}) \equiv U = \gamma/4\pi^2 m^2$\cite{swavefootnote}.

\begin{figure}
\epsfxsize=7.0cm
\epsfysize=12.5cm
\epsffile{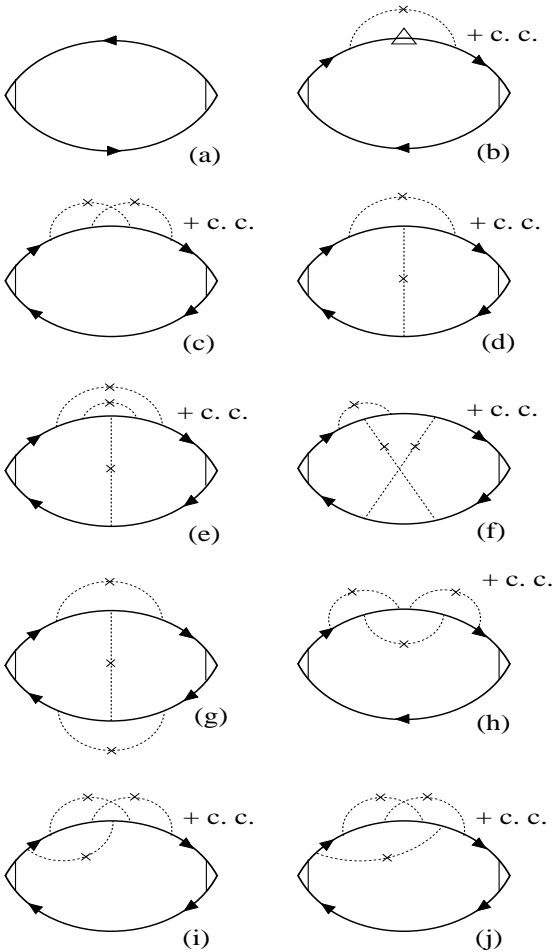}
\vskip 0.5cm
\caption{Simple diagrams that contribute to the terms shown in 
 Eq.\ (\protect\ref{eq:2a}). For each of the diagrams (f)  
 and (j) there is one equivalent symmetric diagram (not shown)
 that also contributes, and the complex conjugates of diagrams
 (b) - (f) and (h) - (j) contribute as well. The `triangulated' line in
 diagram (b) denotes a Green's function, Eq.\ (\protect\ref{eq:4b}), with
 its value at $\gamma=0$ subtracted to avoid double counting. Diagram (a)
 yields $\sigma_B + O(\gamma^2)$ and serves to normalize all other
 contributions. Diagrams (e) - (g) contribute to the $\gamma^{3/2}$ term.}
\label{fig:1}
\end{figure}

\begin{figure}
\epsfxsize=7.0cm
\epsfysize=4.7cm
\epsffile{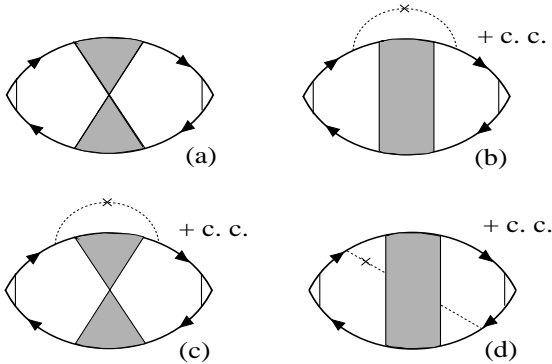}
\vskip 0.5cm
\caption{Infinite resummations that contribute to the terms shown in
 Eq.\ (\protect\ref{eq:2a}). The complex conjugates of diagrams
 (b) - (d) contribute as well.}
\label{fig:2}
\end{figure}

\begin{figure}
\epsfxsize=8.0cm
\epsfysize=3.5cm
\epsffile{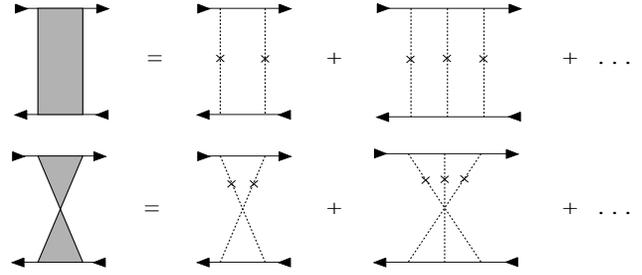}
\vskip 0.5cm
\caption{Definition of the vertices in Fig.\ \protect\ref{fig:2}.}
\label{fig:3}
\end{figure}

The transport coefficient of interest to us, viz. the dynamical conductivity
$\sigma(\omega)$, can be expressed in terms of the Green's functions by
means of the Kubo-Greenwood formula,
\begin{eqnarray}
{\rm Re}\,\sigma(\omega) &=& \frac{e^{2}}{\pi m^{2}} \,{\rm Re}\, 
           \sum_{\bf k,p}\ k_z\ \left\{ {\cal G}^{R}_{\bf k,p}(\omega)\,
{\cal G}^{A}_{\bf p,k}(\omega =0)\right. 
\nonumber\\
 && \left. -\,{\cal G}^{R}_{\bf k,p}(\omega)\,
     {\cal G}^{R}_{\bf p,k}(\omega =0) \right\}_{\rm dis}\ p_z\quad,
\label{eq:5}
\end{eqnarray}
This expression can be used to
systematically expand $\sigma$ in powers of $\gamma$. Such an expansion
has been set up in Ref.\ \onlinecite{KirkpatrickDorfman}, and in 
Ref.\ \onlinecite{Wysolong} all diagrams were identified that contribute
up to and including order $\gamma^2$ in $d=3$. The classification of the
diagrams with respect to the order in $\gamma$ they contribute to
does {\em not} carry over to other values of $d$, but it turns out that
the diagrams that contribute to the terms shown in Eq.\ (\ref{eq:2a})
form a subset of those considered in Ref.\ \onlinecite{Wysolong}. They
are shown in Figs.\ \ref{fig:1} - \ref{fig:3}.

All of these diagrams can be expressed in terms of integrals
over combinations of two functions, $J^{++}$ and $J^{+-}$, that are
convolutions of $G^R$ and $G^A$. At zero frequency they are defined as,
\begin{mathletters}
\label{eqs:6}
\begin{equation}
J^{+\nu}(q) = \int d{\bf k}\ \frac{1}{\epsilon-k^2 + i\gamma}\,
             \frac{1}{\epsilon-({\bf k}-{\bf q})^2 + i\nu\gamma}\quad,
\label{eq:6a}
\end{equation}
with $\nu=\pm$. Doing the integrals yields, in $d=2$,
\begin{equation}
J^{++}(q) = \frac{2\pi}{qw_{++}}\,
      \left[\ln (w_{++} + q) - \ln (w_{++} - q)\right]\quad,
\label{eq:6b}
\end{equation}
\begin{eqnarray}
J^{+-}(q) &=& \frac{-2i\pi}{w_{+-}}\,
 \left[\ln\left(\frac{w_{+-} + 2\gamma -iq^2}{w_{+-} 
                               + 2\gamma +iq^2}\right)\right.
\nonumber\\
 && + \left.\frac{1}{2}\ln\left(\frac{\epsilon - 
      i\gamma}{\epsilon+i\gamma}\right) + i\pi \right]\quad,
\label{eq:6c}
\end{eqnarray}
where,
\begin{equation}
w_{++} = \sqrt{w(q) - 4 i\gamma}\quad,\quad
                w_{+-} = \sqrt{-q^2w(q)+4\gamma^2}\quad,
\label{eq:6d}
\end{equation}
\end{mathletters}%
with $w(q) = q^2 - 4\epsilon$.

The only diagrams that contribute to the
$\gamma^{3/2}$ term in Eq.\ (\ref{eq:2a}) are (e), (f), and (g)
in Fig.\ \ref{fig:1}. In order to demonstrate how
the nonanalyticity arises, let us consider diagram (g) as an example. After
simple manipulations, its contribution to the conductivity can be written,
\begin{equation}
\frac{\sigma^{\rm (1g)}}{\sigma_{B}} = \left(\frac{\gamma}{2\epsilon}\right)^2
  \,\frac{4\epsilon^2}{\pi^8} \int d{\bf q}\ {\rm Re}\,J^{++}(q)\,
                             {\rm Re}[J^{++}(q)]^2\quad.
\label{eq:7}
\end{equation}
Here we have put $\omega=0$, and have kept only the leading contribution
for $\gamma\rightarrow 0$. It is now easy to see how the nonanalyticity
arises. Equation (\ref{eq:6b}) shows that, in the limit $\gamma\rightarrow 0$,
$J^{++}$ contains a singularity of the form $(q-2k_{\rm F})^{-1/2}$. A
$q$-integration over $(J^{++})^3$ thus leads to a $\gamma^{-1/2}$ term. Since
$(J^{++})^3$ first appears in the integrands at order $\gamma^2$, the leading
singularity produced by this mechanism is of the form $\gamma^{3/2}$.
Asymptotic analysis yields the prefactor of the nonanalyticity, which gives
a contribution to the number $\sigma_{3/2}$ in Eq.\ (\ref{eq:2a}) as stated
in Table \ref{tab:1}.
Similarly, an integral over $(J^{++})^2$ produces an $\ln\gamma$, and this
mechanism contributes to the prefactor $\sigma_{1\ln}$ in Eq.\ (\ref{eq:2a}).
Another contribution to $\sigma_{1\ln}$ comes from the `long-wavelength' terms,
which manifest themselves as integrals $\int d{\bf q}\ [J^{+-}(q)]^2$.
Such integrals arise from Fig.\ \ref{fig:1} (d) (in (c) they cancel), and
Fig.\ \ref{fig:2} (a). The fact that in $d=2$ two powers of $J^{+-}$
are sufficient to produce a logarithm accounts for the fact that this
nonanalyticity appears already at linear order in $\gamma$. Finally,
some of the diagrams discussed so far, and all of the remaining ones,
contribute to the analytic term at order $\gamma$. The various contributions
are listed in Table \ref{tab:1}.

\begin{table}
\caption{Contributions of diagrams as shown in Figs.\ \protect\ref{fig:1} and
 \protect\ref{fig:2} to the coefficients in Eq.\ (\protect\ref{eq:2a}).
 No entry means that the diagram does not contribute for structural
 reasons, while a zero indicates no contribution due to cancellations.}
\begin{center}
\begin{tabular}{cccc}
diagram         & $\sigma_{1\ln}$ & $\sigma_{1}$  & $\sigma_{3/2}$ \\ 
\hline
1\ (b)          &                 & $2/\pi$       &                \\
1\ (c)          &                 & $0$           &                \\
1\ (d)          &  $-2/\pi$       &$-(8/\pi)\ln 2$&                \\
1\ (e)          &                 &               &  $-\sqrt{2}/2$ \\
1\ (f)          &                 &               &  $\sqrt{2}$    \\
1\ (g)          &                 &               &  $\sqrt{2}/4$  \\
1\ (h)          &                 &               &  $0$           \\
1\ (i)          &                 &               &  $0$           \\
1\ (j)          &                 &               &  $0$           \\
2\ (a)          &  $2/\pi$        &$(4/\pi)\ln 2$ &                \\
2\ (b)+(c)+(d)  &                 &$(2/\pi)\ln 2$ &                \\
\end{tabular}
\end{center}
\label{tab:1}
\end{table}

The infinite resummation denoted by diagram (a) in Fig.\ \ref{fig:2}
plays a special role in our perturbation theory, and 
deserves some discussion. This `crossed ladder' resummation is the only
diagram where the zero-frequency limit needs to be handled with some
care, since it is related to the so-called weak-localization anomaly,
i.e. the fact that the conductivity of disordered noninteracting electrons
in $d=2$ contains a $\ln\omega\tau$ in perturbation theory, and that the
true zero-frequency value of $\sigma$ is zero\cite{localization}.
Taken at face value, the diagram is finite at $\omega=0$, since
the resummation leads to a structure
$1/(1-\gamma J^{+-}(q)/\pi^2)$. Expanding $J^{+-}$ in powers of $\gamma$
leads to a diffusion pole, i.e. a $1/q^2$ singularity, at lowest order,
but the subleading contribution of $O(1)$ to $J^{+-}(q=0)$ seems to protect the
singularity. 
This is misleading, however, since
it is well known that the `crossed ladder' is an approximation to an
exact vertex part $\Lambda(q,\omega)$ that has an {\em exact} diffusion
pole, $\Lambda(q,\omega) \sim 1/(-i\omega + q^2/D)$, with $D$ the
diffusion coefficient\cite{VW}. Indeed one can show that there exist
classes of diagrams that cancel the mass in the simple crossed ladder,
order by order in perturbation theory\cite{TRK}. Although formally of
higher order in $\gamma$, these contributions lead to the logarithmic
singularity stemming from $\Lambda$ being protected only by a finite
frequency, and {\em not} by $\gamma$. On a more formal level, strict
perturbation theory in powers of $\gamma$ violates
a Ward identity that reflects particle number conservation in the
presence of time reversal invariance. By choosing a self energy that
is related to the crossed ladder vertex correction by means of this
Ward identity, one can construct a conserving approximation for
$\Lambda$ which has the diffusion pole built in.
Replacing the small-$q$ part of
the crossed ladder by
the appropriate exact diffusion pole then leads to the term $\ln\omega\tau$
with a prefactor as first reported by Gorkov et al., 
Ref.\ \onlinecite{localization}, and made rigorous by Kirkpatrick and
Dorfman\cite{KirkpatrickDorfman}, and shown
in Eq.\ (\ref{eq:2a}). Apart from this, the first term in the
infinite crossed ladder resummation also contributes to the coefficients
$\sigma_1$ and $\sigma_{1\ln}$, see Table \ref{tab:1}.
 
We have ascertained that no other diagrams contribute to the terms we
are considering. For simple diagrams,
it is easy to show, by a combination of the diagram rules with power counting,
that diagrams with more than three impurity lines cannot contribute. For
the infinite resummations, the same type of argument shows that dressing
the diagrams shown in Fig.\ \ref{fig:2} by additional impurity lines leads
to terms that vanish faster than $\gamma^{3/2}$. Finally, arguments analogous
to those employed before in $d=3$\cite{Wysolong} show that one need not
consider diagrams with more than one ladder or crossed ladder resummation. 

We also mention that the temperature dependent conductivity is easily
obtained from our $T=0$ result by a convolution with the derivative of a
Fermi function\cite{Greenwood,Wysolong}. At low temperatures, the
$\gamma^{3/2}$ anomaly then gets translated into a $T^{3/2}$ dependence at
fixed scatterer density.
After subtracting out the weak-localization term, this should be observable,
at least in principle, in experiments of the type reported by 
Adams\cite{Adams}.

We gratefully acknowledge helpful discussions with T.R. Kirkpatrick.
This work was supported by the NSF under grant number DMR-95-10185.


\begin{references}
\bibitem{2dfootnote} In a $2-d$ real classical fluid, the static diffusivity
 does not exist due to long-time tail effects, so one also needs to consider
 the frequency dependence. A similar effect occurs in the quantum Lorentz
 model to be studied below.
\bibitem{swavefootnote} In $d=3$, pure s-wave scattering is an excellent
 approximation to the scattering processes in certain 
 systems\cite{Wysolong}. In
 $d=2$ this is not the case, and a more realistic description of the
 scattering process would lead to quantitative changes of our coefficients.
 In particular, $\sigma_1 =0$ may
 hold only for point-like scatterers. However, the
 existence of the $\gamma^{3/2}$ term is independent of this approximation.
\end{references}
\end{document}